\documentclass[preprint,showpacs,eqsecnum,aps]{revtex4}
\usepackage{graphicx}
\usepackage{epsfig}

\begin{document}

\title{Description of the $2\nu\nu\beta\beta$ decay within a fully renormalized pnQRPA approach with a restored gauge symmetry}

\author{C. M. Raduta$^{a)}$, A. A. Raduta$^{a),b)}$ }

\address{$^{a)}$ Department of Theoretical Physics, Institute of Physics and
  Nuclear Engineering, Bucharest, POBox MG6, Romania}

\address{$^{b)}$Academy of Romanian Scientists, 54 Splaiul Independentei, Bucharest 050094, Romania}

\date{\today}

\begin{abstract}
A many body Hamiltonian involving the mean field for a projected spherical single particle basis, the pairing interactions for alike nucleons and the  dipole-dipole proton-neutron interactions in the particle-hole ($ph$) and particle-particle ($pp$) channels is treated by the projected gauge fully renormalized proton-neutron quasiparticle random phase approximation (PGFRpnQRPA) approach.
The resulting wave functions and energies for the mother and the daughter nuclei are used to calculate the $2\nu\beta\beta$ decay rate and the process half life. For illustration,
the formalism is applied for the decay  $^{100}$Mo$ \to$ $^{100}$Ru. The results are in good agreement with the corresponding experimental data. The Ikeda sum rule ($ISR$) is obeyed. The gauge projection makes the $pp$ interaction inefficient.
\end{abstract}
\pacs{23.40.Hc,23.40.-s,21.10.Tg,21.60.Jz, 13.10,+q}
\renewcommand{\theequation}{\arabic{equation}}
\setcounter{equation}{0}
\maketitle
Double beta decay is one of the most exciting topic of nuclear physics since the rate of the process is obtained by
combining formalisms of electroweak interaction with those yielding nuclear matrix elements. Due to this feature
it represents a sensitive test for both collaborating fields. The $2\nu\beta\beta$ process is interesting by its own but is also very attractive because it constitutes  a test for the nuclear matrix elements (m.e.) which are used for the process
of $0\nu\beta\beta$ decay. Discovery of this process may provide an answer for the fundamental question,  whether neutrino is a Mayorana or a Dirac particle.
The subject development is reflected be several review papers 
\cite{PriRo,HaSt,Ver,Suh,Tomo,Fass,Rad1}.
Our contribution described in this letter concerns the $2\nu\beta\beta$ process, which can be viewed as two consecutive and virtual single $\beta^-$ decays. The formalism yielding closest results to the experimental data  is the proton-neutron random phase approximation (pnQRPA) which includes the particle-hole ($ph$) and 
particle-particle ($pp$) as independent two body interactions. The second leg of the $2\nu\beta\beta$ process is very sensitive to changing the relative strength of the later interaction, denoted hereafter by $g_{pp}$.
It is worth mentioning that the two body interaction of $ph$ type is repulsive while that of $pp$ nature is attractive.
Due to this feature there is a critical value for $g_{pp}$ for which the first root of the pnQRPA equation vanishes.
Actually, this is the signal that the pnQRPA approach is no longer valid. Moreover, the $g_{pp}$ value which corresponds to a transition amplitude which agrees with the corresponding experimental data is closed to the mentioned critical value.
That means that the result is not stable to adding corrections to the RPA picture.
The first improvement for the pnQRPA was achieved by one of us (AAR) in Refs.\cite{Rad2,Rad3}, by using a boson expansion (BE) procedure. Later on another procedure showed up, which renormalized the dipole two quasiparticle operators by
replacing the scalar components of their commutators by their average values \cite{Suho1}. 
Such a renormalization is inconsistently achieved since the scattering operators are not renormalized. This lack of consistency was removed in Ref.
\cite{Rad4,Rad5} where a fully renormalized pnQRPA is proposed.

Unfortunately, all higher pnQRPA procedures mentioned above have a common drawback of violating the Ikeda sum rule ($ISR$) by an amount of about 20-30\% \cite{Rad6}. It is believed that such a violation is caused by the gauge symmetry breaking. Consequently,
a method of restoring this symmetry was formulated by the present authors in Ref. \cite{Rad7}.

In this paper the results of Ref.\cite{Rad7} are improved in two respects: a) aiming at providing a unitary description of the process for the situations when the nuclei involved are spherical or deformed, here we use the projected spherical single particle basis defined in Ref.\cite{Rad8} and used for double beta decay in Ref.\cite{Rad9,Rad10}. b) the space of proton-neutron dipole configurations is split in three subspaces, one being associated to the single $\beta^-$, one to the $\beta^+$ process, and one spanned by the unphysical states. A compact expression for the dispersion equation of energies is obtained from the linearized equation of motion of the basic transition operators corresponding to the two coupled processes. The numerical application is made for the $2\nu\beta\beta$ process $^{100}Mo\to ^{100}Ru$. Aiming at a selfcontent presentation, minimal details are necessary.

According to Ref.\cite{Rad8} the projected spherical basis is defined as:
\begin{equation}
\Phi_{nlj}^{IM}(d)={\cal N}_{nlj}^IP_{MI}^I[|nljI\rangle\Psi_g]\equiv
{\cal N}_{nlj}^I\Psi_{nlj}^{IM}(d) ,
\label{phiim}
\end{equation}
where $P^I_{MK}$ denotes the angular momentum projection operator, $|nljm\rangle$ is the spherical
 shell model state and $\Psi_g$ is an axially deformed coherent state describing the ground state of a phenomenological core in terms of quadrupole bosons $b^{\dagger}_{2\mu},b_{2\mu}$:
\begin{equation}
{\Psi}_g=exp[d(b_{20}^+-b_{20})]|0\rangle_b .
\label{psig}
\end{equation}
Here $|0\rangle_b$ denotes the vacuum state for the quadrupole bosons.
The set of functions $\Phi_{nlj}^{IM}(d)$ is an orthogonal single particle basis.
The single particle energies $\epsilon^I_{nlj}$ are obtained by averaging a particle-core Hamiltonian with the corresponding basis states.
In order to keep close to the Nilsson model, where on each $\Omega$ state one can distribute 2 nucleons, here we change the norm of the projected states such that this restriction holds also for each $I$ state:
\begin{equation}
\langle\Phi_{\alpha}^{I M}|\Phi_{\alpha}^{I M}\rangle=1 \Longrightarrow \sum_{M}\langle\Phi_{\alpha}^{IM}|\Phi_{\alpha}^{IM}\rangle=2.
\label{newnorm}
\end{equation} 
Thus, the wave functions used to calculate the m.e. should be multiplied with the statistical factor $\sqrt{2/(2I+1)}$.

We suppose that the states describing the nuclei involved in a $2\nu\beta\beta$ process are described by a many body Hamiltonian which  may be written in the projected spherical basis as:
\begin{eqnarray}
H=&&\sum_{\tau,\alpha,I,M} \frac{2}{2I+1}(\epsilon_{\tau\alpha I}-
\lambda_{\tau\alpha})c^{\dagger}_{\tau\alpha IM}c_{\tau \alpha IM}-\sum_{\tau,\alpha,I,I^{'}}
\frac{G_{\tau}}{4}
P^{\dagger}_{\tau \alpha I}P_{\tau\alpha I'}\nonumber \\
 &+& 2\chi\sum_{pn;p^{'}n^{'};\mu}\beta^-_{\mu}(pn)\beta^+_{-\mu}(p'n')(-)^{\mu}
 -2\chi_1\sum_{pn;p^{'}n^{'};\mu} P^-_{\mu}(pn)P^+_{-\mu}(p'n')(-)^{\mu},
\label{Has}
\end{eqnarray}
where $c^{\dagger}_{\tau\alpha IM}(c_{\tau \alpha IM})$ denotes the creation (annihilation) operator of one nucleon of the type $\tau (=p,n)$ in the state $\Phi^{IM}_{\alpha}$, with $\alpha$ being an abbreviation for the set of quantum numbers $nlj$. The Hamiltonian H contains the mean field term, the pairing interaction for alike nucleons and the Gamow-Teller dipole-dipole interaction in the $ph$ and $pp$ channels, characterized by the strengths $\chi$ and $\chi_1$, respectively.  
Passing to the quasiparticle representation through the Bogoliubov-Valatin transformation:
\begin{equation}
a^{\dagger}_{\tau IM}=U_{\tau I}c^{\dagger}_{\tau IM}-s_{IM}V_{{\tau}I}c_{\tau I-M},
s_{IM}=(-)^{I-M},\;\;\tau=p,n,\;\;
U_{{\tau}I}^2+V_{{\tau}I}^2=1,
\end{equation} 
the first two terms of H are replaced by the independent quasiparticles term, $\sum E_{\tau I}a^{\dagger}_{\tau IM}a_{\tau IM}$, while the $ph$ and $pp$ interactions are expressed in terms of the dipole two $qp$ and the $qp$ density operators:
\begin{eqnarray}
A^{\dag}_{1\mu}(pn)&=&\sum C^{I_p\;I_n\;1}_{m_p\;m_n\;\mu}a^{\dag}_{pI_pm_p}
a^{\dag}_{nI_nm_n},\;\; A_{1\mu}(pn)=\left(A^{\dag}_{1\mu}(pn) \right)^{\dag},
\nonumber\\
B^{\dag}_{1\mu}(pn)&=&\sum C^{I_p\;I_n\;1}_{m_p\;-m_n\;\mu}a^{\dag}_{pj_pm_p}
a_{nI_nm_n}(-)^{I_n-m_n},\;\; B_{1\mu}(pn)=\left(B^{\dag}_{1\mu}(pn) \right)^{\dag}.
\end{eqnarray}
In Ref.\cite{Rad4}, we showed that all these operators can be renormalized as suggested by the commutation equations:
\begin{eqnarray}
\left[A_{1\mu}(k),A^{\dagger}_{1\mu'}(k')\right] &\approx &\delta_{k,k'}\delta_{\mu,\mu'}
\left[1-\frac{{\hat N}_n}{{\hat I}^2_n}-\frac{{\hat N}_p}{{\hat I}^2_p}\right],
\nonumber\\
 \left[B^{\dagger}_{1\mu}(k),A^{\dagger}_{1\mu'}(k')\right] &\approx &
\left[B^{\dagger}_{1\mu}(k),A_{1\mu'}(k')\right] \approx 0,
\nonumber\\
 \left[B_{1\mu}(k),B^{\dagger}_{1\mu'}(k')\right] &\approx &\delta_{k,k'}\delta_{\mu,\mu'}
\left[\frac{{\hat N}_n}{{\hat I}^2_n}-\frac{{\hat N}_p}{{\hat I}^2_p}\right]
,\;k=(I_p,I_n).
\label{comm2}
\end{eqnarray}
Indeed, denoting by $C^{(1)}_{I_p,I_n}$ and $C^{(2)}_{I_p,I_n}$ the averages of the right hand sides
of (\ref{comm2}) with the renormalized RPA vacuum state, the renormalized operators defined as
$\bar{A}_{1\mu}(k)=\frac{1}{\sqrt{C^{(1)}_k}}A_{1\mu},\;\bar{B}_{1\mu}(k)=\frac{1}{\sqrt{|C^{(2)}_k|}}
B_{1\mu}$,
obey boson like commutation relations:
\begin{eqnarray}
\left[\bar{A}_{1\mu}(k), \bar{A}^{\dagger}_{1\mu'}(k')\right ] &=&\delta_{k,k'}\delta_{\mu,\mu'},
\nonumber\\
\left[\bar{B}_{1\mu}(k), \bar{B}^{\dagger}_{1\mu'}(k')\right ] &=&\delta_{k,k'}\delta_{\mu,\mu'}f_k,
\;\;f_k=sign(C^{(2)}_k ).
\end{eqnarray}
Further, these operators are used to define the phonon operator:
\begin{equation}
C^{\dagger}_{1\mu}=\sum_{k}\left[X(k)\bar{A}^{\dagger}_{1\mu}(k)+
Z(k)\bar{D}^{\dagger}_{1\mu}(k)- Y(k)\bar{A}_{1-\mu}(k)(-)^{1-\mu} -
W(k)\bar {D}_{1-\mu}(k)(-)^{1-\mu}\right],
\end{equation}
where $ \bar{D}^{\dagger}_{1\mu}(k)$ is equal to $\bar{B}^{\dagger}_{1\mu'}(k')$ or $\bar{B}_{1\mu}(k)$ 
depending on whether $f_k$ is + or -. The phonon amplitudes are determined by
the equations supplied by the operator equations:
\begin{equation}
\left[H,C^{\dagger}_{1\mu}\right]=\omega C^{\dagger}_{1\mu}\;\;\left[C_{1\mu},C^{\dagger}_{1\mu'}\right]=\delta_{\mu\mu'}.
\end{equation}
Interesting properties for these equations and their solutions are discussed in our previous publications \cite{Rad4,Rad5}. Here we mention one of these features. The renormalized ground state is a superposition of components describing the neighboring nuclei $(N-1,Z+1),(N+1,Z-1),(N+1Z+1),(N-1,Z-1)$. The first two components conserve the total number of nucleons (N+Z) but violates the third component of isospin, $T_3$. By contrast, the last two components  violates the total number of nucleons but preserve $T_3$. Actually, the last two components contribute to the violation of the $ISR$. One can construct linear combinations of the basic operators $A^{\dagger},A, B^{\dagger},B$ 
which excite the nucleus $(N,Z)$ to the nuclei $(N-1,Z+1),(N+1,Z-1),(N+1,Z+1),(N-1,Z-1)$, respectively. These operators are:
\begin{eqnarray}
{\cal A}^{\dag}_{1\mu}(pn)&=&U_pV_nA^{\dag}_{1\mu}(pn)+U_nV_pA_{1,-\mu}(pn)(-)^{1-\mu}+
U_pU_nB^{\dag}_{1\mu}(pn)-V_pV_nB_{1,-\mu}(pn)(-)^{1-\mu},\nonumber\\
{\cal A}_{1\mu}(pn)&=&U_pV_nA_{1\mu}(pn)+U_nV_pA^{\dag}_{1,-\mu}(pn)(-)^{1-\mu}+
U_pU_nB_{1\mu}(pn)-V_pV_nB^{\dag}_{1,-\mu}(pn)(-)^{1-\mu},
\nonumber\\
{\bf{\bf A}}^{\dag}_{1\mu}(pn)&=&U_pU_nA^{\dag}_{1\mu}(pn)-V_pV_nA_{1,-\mu}(pn)(-)^{1-\mu}-
U_pV_nB^{\dag}_{1\mu}(pn)-V_pU_nB_{1,-\mu}(pn)(-)^{1-\mu},
\nonumber\\
{\bf{\bf A}}_{1\mu}(pn)&=&U_pU_nA_{1\mu}(pn)-V_pV_nA^{\dag}_{1,-\mu}(pn)(-)^{1-\mu}-
U_pV_nB_{1\mu}(pn)-V_pU_nB^{\dag}_{1,-\mu}(pn)(-)^{1-\mu}.\nonumber
\end{eqnarray}
Indeed, in the particle representation these operators have the expressions:
\begin{eqnarray}
{\cal A}^{\dag}_{1\mu}(pn)&=&-\left[c^{\dag}_pc_{\widetilde{n}}\right]_{1\mu},\;\;
{\cal A}_{1\mu}(pn)=-\left[c^{\dag}_pc_{\widetilde{n}}\right]^{\dag}_{1\mu},
\nonumber\\
{\bf{\bf A}}^{\dag}_{1\mu}(pn)&=&\left[c^{\dag}_pc^{\dag}_n\right]_{1\mu},\;\;
{\bf{\bf A}}_{1\mu}(pn)=\left[c^{\dag}_pc^{\dag}_n\right]^{\dag}_{1\mu}.\nonumber
\end{eqnarray}
In terms of the new operators the many body Hamiltonian is:
\begin{eqnarray}
H&=&\sum_{\tau jm}E_{\tau j}a^{\dag}_{\tau jm}a_{\tau jm}+
2\chi\sum_{pn,p'n';\mu}\sigma_{pn;p'n'}
{\cal A}^{\dag}_{1\mu}(pn){\cal A}_{1\mu}(p'n')  -
2\chi_1\sum_{pn,p'n';\mu}\sigma_{pn;p'n'}
{\bf {\bf A}}^{\dag}_{1\mu}(pn){\bf {\bf A}}_{1\mu}(p'n'),\nonumber\\
\sigma_{pn;p'n'}&=&\frac{2}{3{\hat I}_n{\hat I}_{n^{\prime}}}\langle I_p||\sigma ||I_n\rangle
\langle I_{p'}||\sigma ||I_{n'}\rangle .
\end{eqnarray}
Here $E_{\tau I}$ denotes the quasiparticle energy. Since we are interested in describing the harmonic modes which preserve the total number of nucleons, we ignore the $\chi_1$ term. Indeed, this term defines a deuteron type excitation, which will be studied in a separate publication.
The equations of motion of the operators defining the phonon operator are determined  by the commutation relations:
\begin{equation}
\left[{\cal A}_{1\mu}(pn),{\cal A}^{\dag}_{1\mu '}(p'n')\right]\approx
\delta_{\mu,\mu'}\delta_{j_p,j_{p'}}\delta_{j_n,j_{n'}}
\left[U_p^2-U_n^2+\frac{U_n^2-V_n^2}{{\hat j}_n^2}{\hat N}_n-
\frac{U_p^2-V_p^2}{{\hat j}_p^2}{\hat N}_p\right].
\end{equation}
The average of the r.h. side of this equation with the $PGFRpnQRPA$ vacuum state
is denoted by:
\begin{equation}
D_1(pn)=U_p^2-U_n^2+\frac{1}{2I_n+1}(U_n^2-V_n^2)\langle{\hat N}_n\rangle -
\frac{1}{2I_p+1}(U_p^2-V_p^2)\langle{\hat N}_p\rangle .
\label{D1pn}
\end{equation}
The equations of motion show that the two $qp$ energies are renormalized too:
\begin{equation}
E^{ren}(pn)=E_{p}(U^2_p-V^2_p)+E_n(V^2_n-U^2_n).
\end{equation}
The space of $pn$ dipole states, ${\cal S}$, is written as a sum of three subspaces defined as:
\begin{eqnarray}
{\cal S}_{+}&=&\left\{(p,n)|D_1(pn)> 0,\;\;E^{ren}(pn)> 0,\right\},\;\;  
{\cal S}_{-}=\left\{(p,n)|D_1(pn)< 0,\;\;E^{ren}(pn)< 0,\right\},\nonumber\\  
{\cal S}_{sp}&=&{\cal S}-\left({\cal S}_{+}+{\cal S}_{-}\right).
\end{eqnarray}
In ${\cal S}_{+}$ one defines the renormalized operators:
\begin{equation}
\bar{\cal A}^{\dagger}_{1\mu}(pn)=\frac{1}{\sqrt{D_1(pn)}}{\cal A}^{\dagger}_{1\mu}(pn),\;\; 
\bar{\cal A}_{1\mu}(pn)=\frac{1}{\sqrt{D_1(pn)}}{\cal A}_{1\mu}(pn),\;\;
\end{equation}
while in  ${\cal S}_{-}$ the renormalized operators are:
\begin{equation}
\bar{\cal F}^{\dagger}_{1\mu}(pn)=\frac{1}{\sqrt{|D_1(pn)|}}{\cal A}_{1\mu}(pn),\;\; 
\bar{\cal F}_{1\mu}(pn)=\frac{1}{\sqrt{|D_1(pn)|}}{\cal A}^{\dagger}_{1\mu}(pn).
\end{equation}
Indeed, the operator pairs ${\cal A}_{1\mu},{\cal A}^{\dagger}_{1\mu}$ and
${\cal F}_{1\mu},{\cal F}^{\dagger}_{1\mu}$ satisfy  commutation relations of boson type.
An RPA treatment within ${\cal S}_{sp}$ would yield either vanishing or negative energies. 
The corresponding states are therefore spurious.

The equations of motion for the renormalized operators read:
\begin{eqnarray}
\left[H,\bar{\cal A}^{\dagger}_{1\mu}\right]&=&E^{ren}(pn)\bar{\cal A}^{\dagger}_{1\mu}
+2\chi\sum_{(p_1n_1)\in {\cal S}_+}\sigma^{(1)}_{pn;p_1n_1}\bar{\cal A}^{\dagger}_{1\mu}
\nonumber\\
&+&2\chi\sum_{(p_1n_1)\in {\cal S}_-}\sigma^{(1)}_{pn;p_1n_1}\bar{\cal F}^{\dagger}_{1-\mu}(-1)^{1-\mu},\nonumber\\
\left[H,\bar{\cal F}^{\dagger}_{1\mu}\right]&=&|E^{ren}(pn)|\bar{\cal F}^{\dagger}_{1\mu}
+2\chi\sum_{(p_1n_1)\in {\cal S}_-}\sigma^{(1)}_{pn;p_1n_1}\bar{\cal F}^{\dagger}_{1\mu}
\nonumber\\
&+&2\chi\sum_{(p_1n_1)\in {\cal S}_+}\sigma^{(1)}_{pn;p_1n_1}\bar{\cal A}_{1-\mu}(-1)^{1-\mu},
\nonumber\\
\left[H,\bar{\cal A}_{1\mu}\right]&=&-E^{ren}(pn)\bar{\cal A}_{1\mu}
-2\chi\sum_{(p_1n_1)\in {\cal S}_+}\sigma^{(1)}_{pn;p_1n_1}\bar{\cal A}_{1\mu}
\nonumber\\
&-&2\chi\sum_{(p_1n_1)\in {\cal S}_-}\sigma^{(1)}_{pn;p_1n_1}
\bar{\cal F}^{\dagger}_{1,-\mu}(-1)^{1-\mu},\nonumber
\\
\left[H,\bar{\cal F}_{1\mu}\right]&=&-|E^{ren}(pn)|\bar{\cal F}_{1\mu}
-2\chi\sum_{(p_1n_1)\in {\cal S}_-}\sigma^{(1)}_{pn;p_1n_1}\bar{\cal F}_{1\mu}
\nonumber\\
&-&2\chi\sum_{(p_1n_1)\in {\cal S}_+}\sigma^{(1)}_{pn;p_1n_1}
\bar{\cal A}^{\dagger}_{1,-\mu}(-1)^{1-\mu},
\end{eqnarray}
where:
\begin{equation}
\sigma^{(1)}_{pn;p_1n_1}=\frac{2}{\hat{1}\hat{I}_n}\langle p||\sigma||n\rangle|D_1(pn)|^{1/2}
\frac{2}{\hat{1}\hat{I}_{n_1}}\langle p_1||\sigma||n_1\rangle|D_1(p_1n_1)|^{1/2}\equiv T_{pn}T_{p_1n_1}.
\end{equation}
The phonon operator is defined as:
\begin{equation}
\Gamma^{\dagger}_{1\mu}=\sum_{k}\left[X(k)\bar{A}^{\dagger}_{1\mu}(k)+
Z(k)\bar{F}^{\dagger}_{1\mu}(k)- Y(k)\bar{A}_{1-\mu}(k)(-)^{1-\mu} -
W(k)\bar {F}_{1-\mu}(k)(-)^{1-\mu}
\right],
\end{equation}
with the amplitudes determined by the equations:
\begin{equation}
\left[H,\Gamma^{\dagger}_{1\mu}\right]=\omega \Gamma^{\dagger}_{1\mu},\;
\left[\Gamma_{1\mu},\Gamma^{\dagger}_{1\mu^{'}}\right]=\delta_{\mu,\mu^{'}}.
\end{equation}   
The compatibility condition for the homogeneous system of equations determining the phonon amplitudes yields two dispersion equations for $\omega$:
\begin{equation}
2\chi \left({\cal R}^+_{-}-{\cal R}^-_+\right)=1,\;\; 2\chi\left(-{\cal R}^+_{+}+{\cal R}^-_-\right)=1,
\label{DisEq}
\end{equation}
with
\begin{equation}
{\cal R}^{+}_{\pm}=\sum_{(p_1n_1)\in{\cal S}_+}\frac{T^2_{p_1n_1}}{\omega\pm E^{ren}(p_1n_1)},\;\;
{\cal R}^{-}_{\pm}=\sum_{(p_1n_1)\in{\cal S}_-}\frac{T^2_{p_1n_1}}{\omega\pm |E^{ren}(p_1n_1)|}.
\end{equation}
The phonon amplitudes can be analytically determined:
\begin{eqnarray}
X(pn)&=&2\chi\frac{T^2_{pn}}{\omega-E^{ren}(pn)}C,\;\;\;
W(pn)=-2\chi\frac{T^2_{pn}}{\omega+|E^{ren}(pn)|}C,\;\;\nonumber\\
Z(pn)&=&2\chi\frac{T^2_{pn}}{\omega-|E^{ren}(pn)|}C^{'},\;\;
Y(pn)=-2\chi\frac{T^2_{pn}}{\omega+E^{ren}(pn)}C^{'}.
\end{eqnarray}
The constant factors $C$ and $C^{'}$ have the expressions:
\begin{equation}
C^{-2}=4\chi^2\left[{\cal R}^{+}_{-}-{\cal R}^{-}_+\right],\;
C^{'-2}=4\chi^2\left[{\cal R}^{-}_{-}-{\cal R}^{+}_{+}\right].
\end{equation}
In order to solve Eqs.(\ref{DisEq}) we need to know $D_1(pn)$ and, therefore, the averages of the  $qp$'s number operators, $\hat{N}_p$ and $\hat{N}_n$. These are written first in  particle representation and then the particle number conserving term
is expressed as a linear combination of ${\cal A}^{\dagger}{\cal A}$ and ${\cal F}^{\dagger}{\cal F}$ chosen such that their commutators with ${\cal A}^{\dagger}, {\cal A}$ and ${\cal F}^{\dagger},{\cal F}$ are preserved. The final result is:
\begin{eqnarray}
\langle \hat{N}_p \rangle &=&V^2_p(2I_p+1)+3(U^2_p-V^2_p)
(\sum_{\stackrel{n^{'},k}{(p,n^{'})}\in{{\cal S}_+}}D_1(p,n^{'})(Y_k(p,n^{'}))^2-
\sum_{\stackrel{n^{'},k}{(p,n^{'})}\in{{\cal S}_-}}D_1(p,n^{'})(W_k(p,n^{'}))^2),\nonumber\\
\langle \hat{N}_n \rangle &=&V^2_n(2I_n+1)+3(U^2_n-V^2_n)
(\sum_{\stackrel{p^{'},k}{(p^{'},n)}\in{{\cal S}_+}}D_1(p^{'},n)(Y_k(p^{'},n))^2-
\sum_{\stackrel{p^{'},k}{(p^{'},n)}\in{{\cal S}_-}}D_1(p^{'},n)(W_k(p^{'},n))^2).\nonumber
\label{AvN}
\end{eqnarray}
Eqs. (\ref{DisEq}), (\ref{AvN}) are to be solved iteratively. 
It is worth mentioning that using the quasiparticle representation for the basic operators
${\cal A}^{\dagger}_{1\mu},{\cal F}^{\dagger}_{1\mu},{\cal A}_{1,-\mu}(-1)^{1-\mu},
{\cal F}_{1,-\mu}(-)^{1-\mu}$, one obtains for $\Gamma^{\dagger}_{1\mu}$ an expression which involves the scattering $pn$ operators. Thus, the present approach is indeed the $PGFRpnQRPA$.

The formalism presented above was used to describe the $2\nu\beta\beta$ process.
If the energy carried by leptons in the intermediate state is approximated by
the sum of the rest energy of the emitted electron and half the Q-value of
the double beta decay process
\begin{equation}
\Delta E=\frac{1}{2}Q_{\beta\beta}+m_ec^2,
\label{DeltaE}
\end{equation} 
the reciprocal value of the $2\nu\beta\beta$ half life can be factorized as:
\begin{equation}
(T^{2\nu\beta\beta}_{1/2})^{-1}=F|M_{GT}(0^+_i\rightarrow 0^+_f)|^2,
\label{T1/2}
\end{equation}
where F is an integral on the phase space, independent of the nuclear structure,
 while M$_{GT}$ stands for the Gamow-Teller transition amplitude and has
 the expression :
\begin{equation}
M_{GT}=\sqrt{3}\sum_{k,k'}\frac{_i\langle0||\beta^+_i||1_k\rangle_i 
\mbox{}_i\langle1_k|1_{k'}\rangle_f 
\mbox{}_f\langle1_{k'}||\beta^+_f||0\rangle_f}{E_k+\Delta E+E_{1^+}}.
\label{MGT}
\end{equation}
In the above equation, the denominator consists of three terms: a) $\Delta E$,
which was already defined, b) the average value of the k-th $PDFRpnQRPA$ energy
normalized to the particular value corresponding to k=1,
and c) the experimental energy for the lowest $1^+$ state.
The indices carried by the $\beta^+$ operators indicate that they act in the
space spanned by the $PGFRpnQRPA$ states associated to the initial ($i$) or final
($f$) nucleus. The overlap m.e. of the single phonon states in the initial and
final nuclei respectively, are calculated within $GPFRpnQRPA$.  In Eq.(\ref{MGT}), 
the Rose convention for the reduced m.e. is used \cite{Rose}.

Note that if we restrict the  $pn$ space to ${\cal S}_+$, $M_{GT}$ vanishes due to the second leg of the transition. Indeed, the m.e. associated to the daughter nucleus is of the 
type $_f\langle 0|(c^{\dagger}_nc_p)_{1\mu}(c^{\dagger}_nc_p)_{1\mu}|o\rangle_f$, which is equal to zero due to the Pauli principle restriction.
Also, we remark that the operator $\bar{\cal A}^{\dagger}_{1\mu}$ plays the role of a  $\beta^-$ transition operator,
 while when $\bar{\cal F}^{\dagger}_{1\mu}$ is applied
on the ground state of the daughter nucleus, it induces a $\beta^+$ transition. Therefore, the $2\beta$ decay cannot be described by considering the $\beta^-$ transition alone.

For illustration, we present the results for the transition $^{100}$Mo$\to ^{100}$Ru.
For this case the energy corrections involved in Eq.(\ref{MGT}) are:
\begin{equation}
\Delta E =2.026 MeV,\;\; E_{1^+}=0.0 MeV.
\end{equation}
\begin{figure}
\begin{center}
\includegraphics[width=0.6\textwidth]{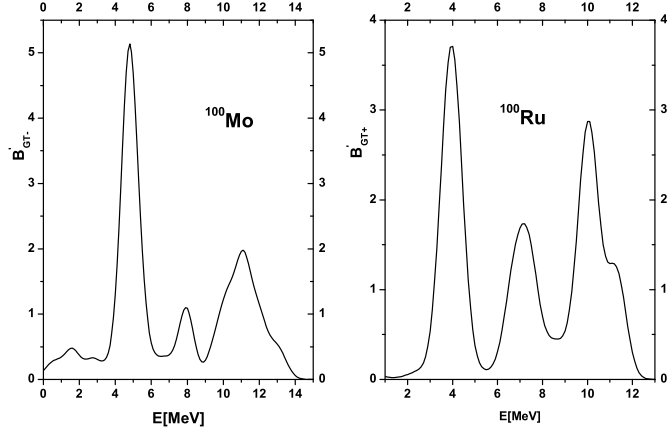}
\end{center}
\caption{The single $\beta^-$ strength for the mother nucleus, $^{100}$Mo (left panel), and the $\beta^+$ strength for the daughter nucleus, $^{100}$Ru, folded by a Gaussian with a width of 0.5MeV, are plotted as functions of the corresponding energies yielded by the present formalism.}    
\end{figure}
\begin{table}
\begin{tabular}{|c|ccccccc|}
\hline
          & d  &k  &  $G_p$[MeV] &$G_n$[MeV]& $ISR$ & log$ft$ & $\chi [MeV]$ \\
\hline
$^{100}$Mo&-1.5&6.1&   0.18      & 0.263     &16.01&  $^{100}$Mo$\stackrel{ \beta^+/EC}{\leftarrow}$$^{100}$Tc& 0.207 \\
          &      &   &             &           &     &4.5~~~~5.29&     \\
\hline
$^{100}$Ru& -0.6&6.1&0.18        &0.25       & 12.01&
$^{100}$Tc$\stackrel{ \beta^-}{\rightarrow}$$^{100}$Ru &0.207 \\
          &       &   &             &          &       &4.66~~~~5.41&    \\
\hline
\end{tabular}
\caption{The deformation parameter d, the pairing interaction strengths for protons ($G_p$) and 
and neutrons ($G_n$) and the GT dipole interaction $\chi$ used in our calculations. We also give
the parameter $k$ relating the quadrupole coordinates and bosons (this is involved in the expression of the single particle energies) as well as the resulting
log$ft$ values characterizing the $\beta^+/EC$ and $\beta_-$ transitions of $^{100}$Tc.
The results for  log$ft$ values, given in the right column, are compared to the experimental data
from the left column.}
\end{table}
The parameters defining the single particle energies are those of the spherical shell model, the deformation parameter $d$ and the parameter $k$ relating the quadrupole coordinate with the quadrupole bosons:
\begin{equation}
\alpha_{2\mu}=\frac{k}{\sqrt{2}}(b^{\dagger}_{2\mu}-(-1)^{\mu}b_{2,-\mu}).
\end{equation}
These are fixed as described in Ref.\cite{Rad10}. The core system is defined by $(Z,N)=(20,20)$. Labeling the states  according to their energies ordering, the single particle space is defined by the indices interval $[11,62]$. The dimensions for the spaces (${\cal S}_+,{\cal S}_-$)
are $(163,1)$ and $(158,5)$ for the mother and daughter nuclei, respectively. The dimension for 
${\cal S}$ is 208 for both the mother and the daughter nuclei.
The strength of the dipole $pn$ was taken to be 
\begin{equation}
\chi=\frac{5.2}{A^{0.7}}MeV .
\end{equation}
This expression was obtained by fitting the positions of the GT resonances in 
$^{40}$Ca, $^{90}$Zr and $^{208}$Pb \cite{Hom}. The value obtained for the $Mo$ and $Ru$ isotopes is that given in Table 1.

Using these input data we calculated the distribution of the $\beta^{\pm}$ strengths with the result shown in Fig.1. The energy intervals where both distributions are large, contribute
significantly to the double beta transition amplitude.

Calculating first the GT transition amplitude and then the Fermi integral with $G_A=1.254$, as in Ref.\cite{Suh},
we obtained the result:
\begin{equation}
|M_{GT}|=0.234,\;\;\; T_{1/2} = 7.89\cdot 10^{18}yr .
\end{equation}
This result should be compared with the experimental results \cite{Elliot,Ejiri}: 
\begin{equation}
T_{1/2}=(8.0\pm 0.16)\cdot 10^{18} yr  ,\;\;
T_{1/2}=(0.115^{+0.03}_{-0.02})\cdot 10^{20} yr .
\end{equation}
Another experimental result concerns the summed strength for the $\beta^-$ transition:
$\sum B_{GT-} =26.69.$
Quenching the theoretical result by a factor $0.6$, as to account for the missing strength, one obtains the value of $28.9$.
The intermediate odd-odd nucleus, $^{100}$Tc, can perform the transition $\beta^+/EC$, feeding
$^{100}$Mo, or the $\beta^-$ transition to $^{100}$Ru. The measured $\log ft$ values for these transitions, are given in Table 1. The corresponding theoretical results are obtained by means of the expression:
\begin{equation}
ft_{\mp}=\frac{6160}{[ {_l}\langle 1_1||\beta^{\pm}||0\rangle_l g_A]^2},\; l=i,f.
\end{equation}
 In order to take account of the effect
of distant states responsible for the "missing strength" in the giant GT resonance \cite{Suh}
we chose $g_A=1.0$.

Summarizing the results of this paper, one may say that restoring the gauge symmetry from the fully renormalized pnQRPA, one obtains a realistic description of the transition rate and moreover the 
$ISR$ is obeyed. As shown in this paper, it seems that there is no need to include the $pp$ interaction in the many body treatment of the process.

{\bf Acknowledgment.} This work was supported by the Romanian Ministry for Education Research Youth and Sport through the CNCSIS projects ID-33/2007 and ID-946/2007.

\end{document}